\documentclass[12pt]{article}
\pdfoutput=1
\newcommand*{\texlive}{2013} % setting tex version; necessary for siunitx_extension
\usepackage[utf8]{inputenc}
\usepackage[T1]{fontenc}
\usepackage[english]{babel} %spell checking, last language is set
\usepackage{xspace}
\usepackage{amsmath,amstext,amssymb}
\usepackage{microtype} %better typesetting
\usepackage[makeroom]{cancel} % cross out parts of equations
\PassOptionsToPackage{usenames,dvipsnames,svgnames,table}{xcolor} % additional colors
\usepackage{xcolor} % additional colors
\usepackage[super]{nth} % 1th, 2nd, 3rd etc.
\usepackage{listings} % code listings
\usepackage{enumitem} % alternative list environments
\setitemize{label=\usebeamerfont*{itemize item}% when using beamer we have to redefine the beamer templates
  \usebeamercolor[fg]{itemize item}
  \usebeamertemplate{itemize item}}
\usepackage{siunitx}
\usepackage{siunitx_extension} % setting stat und sys error
\DeclareSIUnit\Lumi{\per\centi\meter\squared\per\s}
\DeclareSIUnit\timesten{\ensuremath{\times}10}
\DeclareSIUnit\stdDeviation{\ensuremath{\sigma}}
\DeclareSIUnit\GeV{\giga\electronvolt}
\DeclareSIUnit\MeV{\mega\electronvolt}

\usepackage{multirow}
\usepackage{tabulary}
\usepackage{tabularx} %error
\usepackage{colortbl}
\usepackage{chngpage} %temporarily adjust text width -> wide tables
\usepackage{booktabs} % nice top/bottom rules for tables

\PassOptionsToPackage{}{graphicx}         % Grafikeinbindung 1.1
\usepackage{graphicx}         % Grafikeinbindung 1.1
\DeclareGraphicsExtensions{.pdf,.png} % Grafik-Endungen
\usepackage{subfig}
\usepackage{rotating}
\graphicspath{{figures/}{/Users/weidenka/Documents/Promotion/Pictures/}}
\usepackage{tikz}
\tikzstyle{every picture}+=[remember picture]

\usepackage{pdfsync}
\usepackage[heightadjust=all,valign=c]{floatrow}
\usepackage{wrapfig}
\usepackage{blindtext}

\newcommand\pubnumber{\ }
\newcommand\pubdate{\today}

\def\kph{
Institut f\"ur Kernphysik\\
Johannes Gutenberg-Universit\"at Mainz\\ 
Johann-Joachim-Becher-Weg 45\\
D 55128 Mainz, Germany}

\def\Title#1{\begin{center} {\Large #1 } \end{center}}
\def\Author#1{\begin{center}{ \sc #1} \end{center}}
\def\Address#1{\begin{center}{ \it #1} \end{center}}

\newcommand\pubblock{\rightline{\begin{tabular}{l} \pubnumber\\
         \pubdate  \end{tabular}}}
\newenvironment{Abstract}{\begin{quotation}  }{\end{quotation}}
\newenvironment{Presented}{\begin{quotation} \begin{center} 
             PRESENTED AT\end{center}\bigskip 
      \begin{center}\begin{large}}{\end{large}\end{center} \end{quotation}}

\include{babarsym_mod}
\newcommand{\lumi}{\ensuremath{{\cal L}}\xspace}

\newcommand{\Dzero}{\Dz}
\newcommand{\DzeroBar}{\Dzb}

\newcommand{\KK}{\ensuremath{\Kp\Km}\xspace}
\newcommand{\KsKK}{\ensuremath{\KS\KK}\xspace}
\newcommand{\DKsKK}{\ensuremath{\Dz\to\KsKK}\xspace}
\newcommand{\mBC}{\ensuremath{m_{BC}}\xspace}

\newcommand{\bes}{BESIII\xspace}

\PassOptionsToPackage{style=numeric,sorting=none,maxcitenames=2,url=false,isbn=false,doi=false,eprint=false,backend=biber}{biblatex}
\usepackage{biblatex}
\usepackage{csquotes}
\begin{document}
\begin{titlepage}
\pubblock
%{\ensuremath{\cal BR}\xspace}
\vfill
\Title{\huge Observation of SCS decay $D^{+,0}\to\omega\pi$ and branching fraction measurement of $\Dz\to\KsKK$} % Talk title
\vfill
%\Author{Alexey A. Petrov\support}
\Author{Peter Weidenkaff}
% put in address(es) defined above
\Address{\kph}
\vfill
\begin{Abstract}
Using a data set of 2.92 \invfb of \epem collisions at the $\psiprpr$ mass accumulated with the \bes experiment we present preliminary results from our study of the singly Cabibbo-suppressed decays $D\to\omega\pi$ and the decay of \DKsKK. 
   The decay $\Dp\to\omega\pi^+$ is observed for the first time with a significance of 5.4$\sigma$ and we find evidence of 4.1$\sigma$ for the decay $\Dz\to\omega\pi^0$. As a cross-check the branching fraction $D\to\eta\pi$ is measured and is found to be compatible with the current PDG value. 
The branching fraction of the decay \DKsKK is measured in an untagged analysis with 11743$\pm$113 signal events and is found to be $(4.622\pm0.045(stat.)\pm0.181(sys.))\times 10^{-3}$. This is compatible with previous measurements but with significant improved precision.
\end{Abstract}
\vfill
\begin{Presented}
The 7th International Workshop on Charm Physics (CHARM 2015)\\
Detroit, MI, 18-22 May, 2015
\end{Presented}
\vfill
\end{titlepage}
\def\thefootnote{\fnsymbol{footnote}}
\setcounter{footnote}{0}
%

%%%%%%%%%%%%%%%%%%%%%%%%%%%%%%%%%%%%%%%%%%%%%%%%%%%%%%%%%%%%%%%%%%%%%%%%%%%%%
\section{Introduction}
%%%%%%%%%%%%%%%%%%%%%%%%%%%%%%%%%%%%%%%%%%%%%%%%%%%%%%%%%%%%%%%%%%%%%%%%%%%%%
We present two measurements of \D meson branching fractions. Both analyses aim to determine branching fractions precisely and are therefore relevant to improve theoretical predictions of other branching fractions and/or of the \Dz mixing parameters. The search for the decays $\D\to\omega\pi$ and its comparison with theoretical predictions furthermore provides an insight to SU(3) symmetry in \D decays. The analysis of the decay \DKsKK is a step towards a strong phase determination in this channel, which in turn is important in the determination of the CKM angle $\gamma$ via the GGSZ method\cite{Giri:2003ty} in $\Bp\to\Dz h^+$ decays. 

\bes\ is a 4$\pi$ detector with a geometrical acceptance of 93\% and consists of the following components. The momentum and energy loss of charged tracks are measured in a small-cell helium based multilayer drift chamber in a 1T magnetic field. The relative momentum resolution for a \SI{1}{\GeV} track is \SI{0.5}{\percent}, and its energy loss is measured with a precision of \SI{6}{\percent}. The chamber has a radius of \SI{81}{\centi\meter} and is surrounded by a time of flight system built of two layers of plastic scintillator which is capable of measuring the flight time of particles with an accuracy of 80ps in the barrel and 110ps in the end caps. This provides a K$\pi$ separation of 2$\sigma$ for a \SI{0.9}{\GeV} track. Around the time-of-flight system, 6240 CsI(Tl) Crystals measure the energy of electromagnetic showers with a relative resolution of 2.5\%$/\sqrt{E}$ and their position with \SI{0.6}{\centi\meter}/$\sqrt{E}$. Finally, surrounding the superconducting coil of the magnet are 9 layers of resistive plate chambers for muon identification. Further details can be found in \cite{Ablikim:2009aa}.

\bes has collected a large data sample at $\sqrt{s}=\SI{3.773}{\GeV}$ in \epem collisions with an integrated luminosity of \SI{2.92}{\invfb}. At this energy pairs of charged and neutral \D mesons are produced by the decay of the $\psiprpr$ in a quantum-correlated state. Since the additional phase space doesn't allow for another hadron the sample provides a very clean environment to study \D decays.

We present preliminary results for observation of the singly Cabibbo-suppressed decay $D\to\omega\pi$ and the branching fraction measurement of \DKsKK. 
%%%%%%%%%%%%%%%%%%%%%%%%%%%%%%%%%%%%%%%%%%%%%%%%%%%%%%%%%%%%%%%%%%%%%%%%%%%%%
\section{Observation of the SCS decay $D^{+,0}\to\omega\pi$}
%%%%%%%%%%%%%%%%%%%%%%%%%%%%%%%%%%%%%%%%%%%%%%%%%%%%%%%%%%%%%%%%%%%%%%%%%%%%%
\begin{figure}[h]
   \centering
   \includegraphics[width=0.7\textwidth]{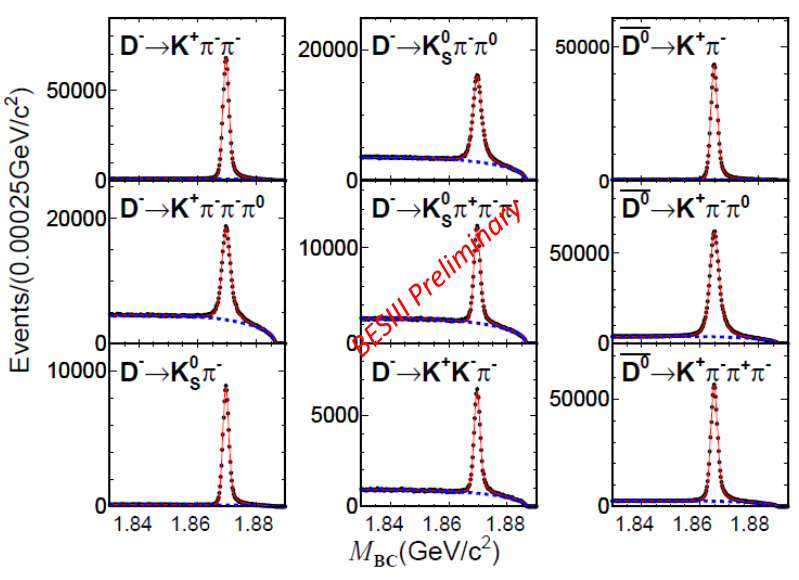}
   \caption{Beam-constraint mass distributions for all tag modes.}
   \label{fig:omegaPi:tag}
\end{figure}
The precise measurement of singly Cabibbo-suppressed decays is challenging since usually statistics are low and background is high. Therefore the clean environment of \D decays at the $\psiprpr$ is ideal to search for and study these decays. The decays of neutral and charged \D mesons to the final state $\omega\pi$ has not been observed yet, but a theoretical calculation exists that predicts the decay at a level of \SI{1}{\timesten\tothe{-4}}\cite{Cheng:2010ry}. CLEO-c failed in a previous analysis to reach that precision and provided a consistent upper limit of \SI{3.0}{\timesten\tothe{-4}} and \SI{2.26}{\timesten\tothe{-4}} @\SI{90}{\percent} C.L. (including \BR($\omega\to\pi^+\pi^-\pi^0$)) for charged and neutral \D decays respectively\cite{Rubin:2005py}. 

With its larger statistics ($\sim 3\times$ CLEO-c), \bes is able to reach the precision of the theoretical prediction. As a cross-check we also extract the branching fractions $\Dp\to\eta\pi^+$ and $\Dz\to\eta\pi^0$.
%-----------------------------------------------------------------------------
\subsection{Reconstruction and selection}
%-----------------------------------------------------------------------------
\begin{wrapfigure}[16]{r}{0.55\textwidth}
   \vspace{-1.2cm}
   \subfloat[$\Dp\to\eta\pi^+$]{
	  \includegraphics[width=0.45\textwidth]{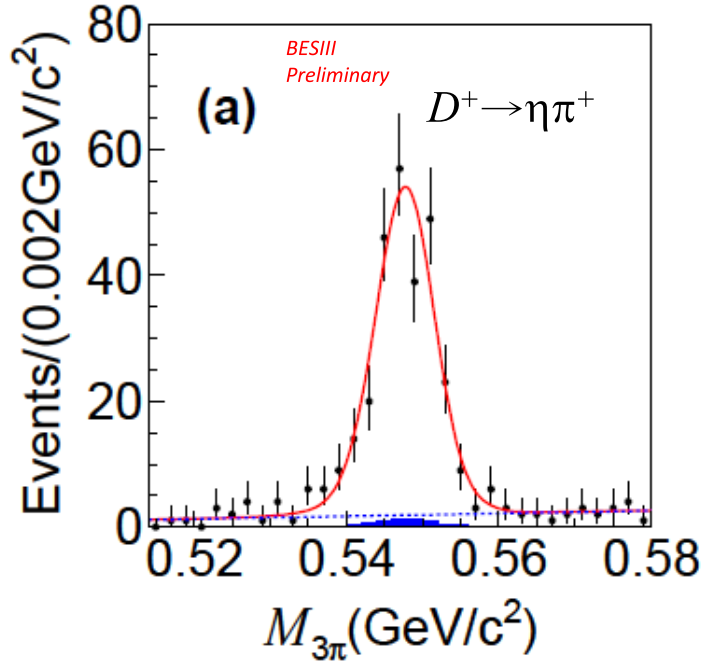}
   }
   \subfloat[$\Dz\to\eta\pi^0$]{
	  \includegraphics[width=0.45\textwidth]{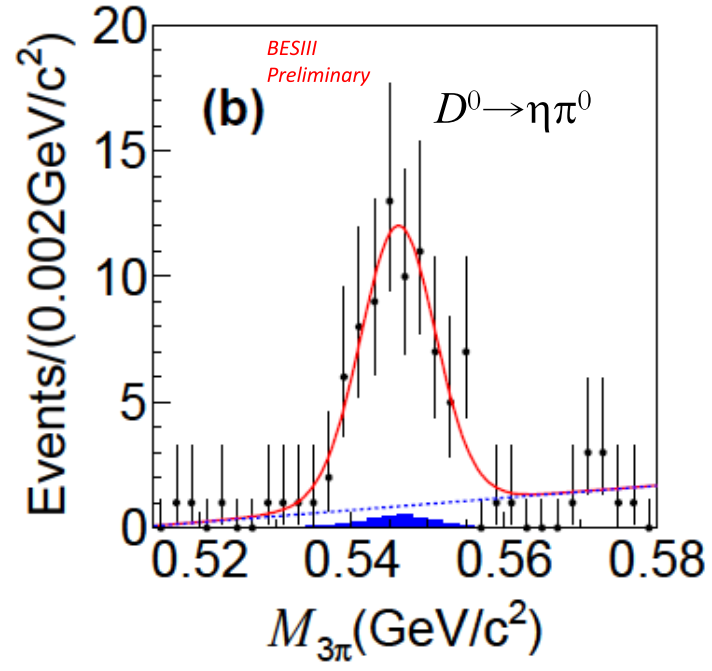}
   }
   \qquad
   \subfloat[$\Dp\to\omega\pi^+$]{
	  \includegraphics[width=0.45\textwidth]{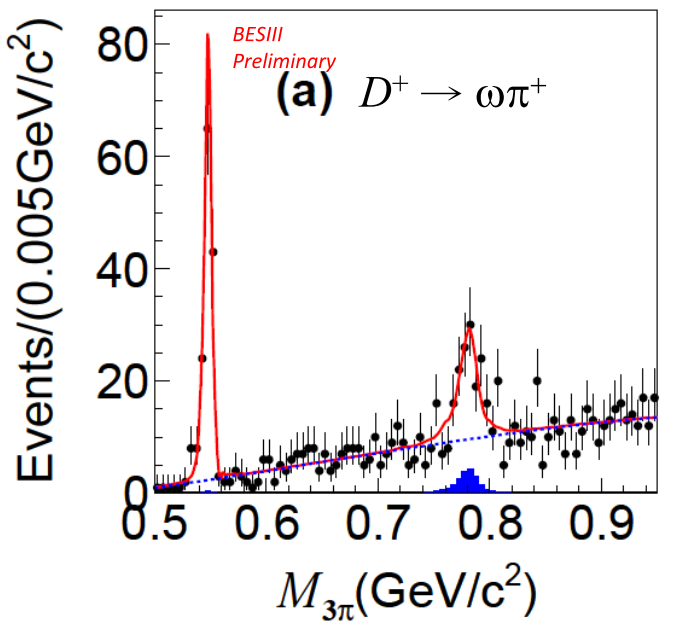}
   }
   \subfloat[$\Dz\to\omega\pi^0$]{
	  \includegraphics[width=0.45\textwidth]{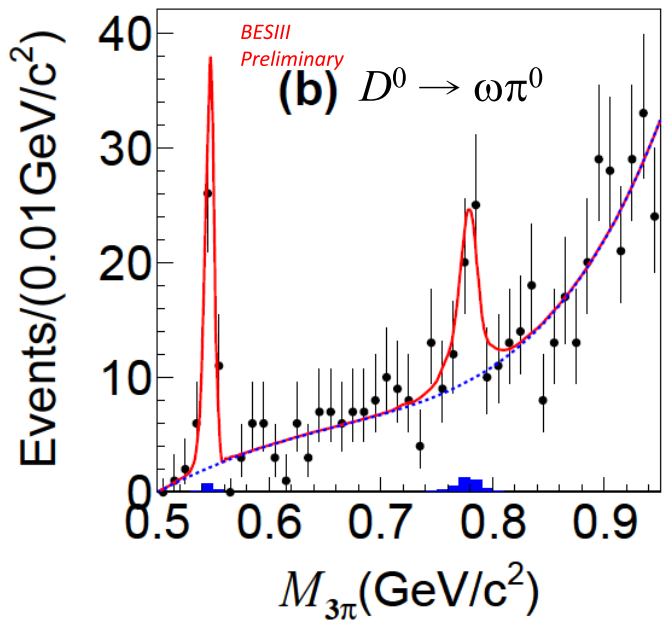}
   }
   \caption{Invariant mass distribution $\pi^+\pi^-\pi^0$.}
   \label{fig:omegaPi:invmassOmega}
  
\end{wrapfigure}
We measure the branching fraction using the so-called double-tag method, which was originally developed by MARKIII\cite{Baltrusaitis:1985iw}. We reconstruct one \D meson in a generic way using a set of decay modes with high branching fractions and low background contamination. We use 6 different modes for the charged \D decay and 3 for the neutral decay. The reconstructed candidates are required to have an energy compatible with the beam energy within approximately 3$\sigma$. If multiple candidates exist the candidate with an energy closest to the beam energy is selected. 

The beam-constraint mass distributions \mBC\footnote{Beam-constraint mass is defined as $\mBC^2=E_{\text{beam}}^2-p_D^2$. With the reconstructed \D momentum $p_D$ and the beam energy $E_{\text{beam}}$.} for all tag modes are shown in Fig.\ref{fig:omegaPi:tag}. From a fit to these distributions with an ARGUS\cite{Albrecht:1994tb} background function and a signal shape that includes effects from ISR, the $\psiprpr$ line shape and detector resolution, we obtain \num{1462041(1359)} and \num{2234741(2425)} tag candidates for the charged and neutral \D decays respectively.

\begin{wrapfloat}{table}[9]{r}{0.5\textwidth}
   \centering
   \resizebox{\textwidth}{!}{
	  \begin{tabular}{c|c|c|c}
		 \toprule
		 & N & N$^{bkg}$ & N$^{obs}_{sig}$\\
		 \midrule
		 $\Dp\to\omega\pi^+$ & \num{98(15)} & \num{22(4)}&\num{76(16)} \\
		 $\Dz\to\omega\pi^0$ & \num{40(11)} &\num{4(8)} &\num{36(14)} \\
		 $\Dp\to\eta\pi^+$ &\num{262(17)} &\num{6(2)} &\num{256(18)} \\
		 $\Dz\to\eta\pi^0$ &\num{71(9)} &\num{3(2)} &\num{68(10)} \\
		 \bottomrule
	  \end{tabular}
   }
   \caption{Signal and background yields.}
   \label{tab:omegaPi:yields}
\end{wrapfloat}
In events in which a tag candidate is found we search for the final states $D^{+}\to(\pi^+\pi^-\pi^0)_{\omega/\eta}\pi^{+}$ and $D^{0}\to(\pi^+\pi^-\pi^0)_{\omega/\eta}\pi^{0}$. Again we select the candidate with the energy closest to the beam energy if multiple candidates exist.
  Two combinations are possible the assign the $\pi^+/\pi^0$ and the wrong combination is almost completely excluded by a requirement on the invariant $3\pi$ mass. The double tag technique highly suppresses background from continuum background (\qqbar). To also suppress the remaining \DD background we require that the helicity H$_\omega$\footnote{The helicity H$_{\omega}$ is defined as the angle between the $\omega$ decay plane and the direction of the \D meson in the $\omega$ rest frame.}  of the $\omega$ is larger 0.54(\Dp) and 0.51(\Dz). Further we apply a $\KS$ veto to suppress background from D$^{+,0}\to\KS\pi^+\pi^{0,-}$. A 2D signal region in the beam-constraint mass of tag and signal decay is defined.
The $(\pi^+\pi^-\pi^0)_{\omega/\eta}$ invariant mass distribution is shown in Fig.\ref{fig:omegaPi:invmassOmega}(c)(d).

%-----------------------------------------------------------------------------
\subsection{Background and signal yield}
%-----------------------------------------------------------------------------
\begin{wrapfloat}{table}[14]{r}{0.5\textwidth}
   \vspace{-0.5cm}
   \centering
   \resizebox{\textwidth}{!}{
	  \begin{tabular}{c|c|c|c|c}
		 \toprule
		 Source & $\omega\pi^\pm$ & $\omega\pi^0$ & $\eta\pi^\pm$ & $\eta\pi^0$ \\
		 \midrule
		 $\pi^\pm$ tracking & 3.0 & 2.0 & 3.0 & 2.0 \\
		 $\pi^\pm$ PID & 1.5 & 1.0 & 1.5 & 1.0 \\
		 $\pi^0$ reconstruction & 1.0 & 2.0 & 1.0 & 2.0 \\

		 2D $M_\mathrm{BC}$ window & 0.1 & 0.2 & 0.1 & 0.2 \\
		 $\Delta E$ requirement & 0.5 & 1.6 & 0.5 & 1.6 \\
		 $|H_{\omega}|$ requirement & 3.4 & 3.4 & -- & -- \\
		 $K^0_S$ veto & 0.8 & 0.8 & -- & -- \\

		 Sideband regions & 0.5 & 6.7 & 0.0 & 0.5 \\

		 Signal resolution \& shape & 0.9 & 0.9 & 4.3 & 5.4 \\
		 Background shape & 3.3 & 2.0 & 2.0 & 3.2 \\
		 Fit range & 0.6 & 1.9 & 0.8 & 1.1 \\

		 $\mathcal{B}(\omega(\eta)\rightarrow\pi^+\pi^-\pi^0)$ & 0.8 & 0.8 & 1.2 & 1.2 \\
		 \midrule
		 Overall & 6.1 & 8.8 & 6.1 & 7.3 \\
		 \bottomrule
	  \end{tabular}
   }
   \caption{Systematic uncertainties.}
   \label{tab:omegaPi:systematics}
   %\makeatletter
   %\def\@captype{figure}
   %\makeatother
   %\includegraphics[width=\textwidth]{OmegaPi/helicity-omegaPi0.png}
   %\caption{Helicity distribution $\Dz\to\omega\pi^0$.}
   %\label{fig:omegaPi:helicity}
\end{wrapfloat}
The signal yield is extracted from the 3$\pi$ invariant mass. The $\omega/\eta$ signal shape is taken from MC and convoluted with a Gaussian to take differences in resolution between data and MC into account. In case of the $\eta$ peak the width is a fit parameter, and for the $\omega$ we use the $\eta$ width scaled with a factor taken from MC. The combinatorial background is described by polynomials. The 'raw' yield N$_{\omega/\eta}$ includes a small component of peaking background from the continuum process $\epem\to(\omega/\eta)+(n\pi)$. We extrapolate events from sideband regions to the signal region and subtract the number of background events to obtain the number of signal decays N$_{\text{sig}}^{\text{obs}}$. The yields are summarized in Tab.\ref{tab:omegaPi:yields}.

%-----------------------------------------------------------------------------
\subsection{Systematics and results}
%-----------------------------------------------------------------------------
\begin{wrapfloat}{figure}{r}{0.5\textwidth}
   \centering
   \includegraphics[width=\textwidth]{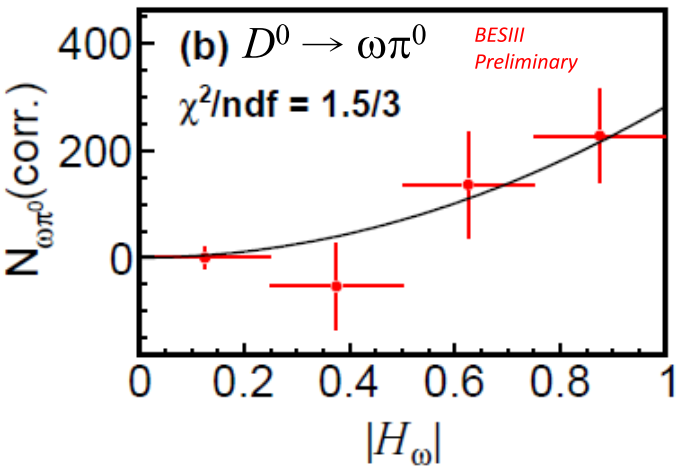}
   \caption{Helicity distribution $\Dz\to\omega\pi^0$.}
   \label{fig:omegaPi:helicity}
   \vspace{-0.3cm}
\end{wrapfloat}
The major source of systematic uncertainties arise from differences between data and MC. The overview of all contributions is shown in Tab.\ref{tab:omegaPi:systematics}. The main contributions come from charged track reconstruction as well as from the requirement on the $\omega$ helicity. The helicity distribution is shown in Fig.\ref{fig:omegaPi:helicity}. The distribution for data follows the expected distribution of the P$\to$VP decay ($\sim\cos^2\theta$). Further significant contributions come from signal and background shapes.

The resulting preliminary branching fractions are listed in Tab.\ref{tab:omegaPi:results}. We are able to observe the decay of charged \D mesons to the final state $\omega\pi^+$ with a significance of 5.4$\sigma$ and we find evidence for the neutral \D decay to $\omega\pi^0$ at the 4.1$\sigma$ level. As a cross-check the branching fractions $D\to\eta\pi$ are also measured for the neutral and charged \D decay. The results are in good agreement with the current PDG\cite{Agashe:2014kda} values.

\begin{table}
   \centering
   \resizebox{\textwidth}{!}{
	  \begin{tabular}{c|c|c}
		 Decay mode & This work & Previous measurements\cite{Aubert:2005sm} \\
		 \hline
		 $\Dp\to\omega\pi^+$ & \SIerrs{2.74}{0.58}{0.17}{\timesten\tothe{-4}} & < \SI{3.4}{\timesten\tothe{-4} @\SI{90}{\percent}}C.L.\\
		 $\Dz\to\omega\pi^0$ & \SIerrs{1.05}{0.41}{0.09}{\timesten\tothe{-4}} & < \SI{2.6}{\timesten\tothe{-4} @\SI{90}{\percent}}C.L.\\
		 \hline
		 $\Dp\to\eta\pi^+$ & \SIerrs{3.13}{0.22}{0.19}{\timesten\tothe{-3}} & \SI{3.53(21)}{\timesten\tothe{-3}}\\
		 $\Dz\to\eta\pi^0$ & \SIerrs{0.67}{0.10}{0.05}{\timesten\tothe{-3}} & \SI{0.68(7)}{\timesten\tothe{-3}}\\
		 \hline
	  \end{tabular}
   }
   \caption{Preliminary results for the branching fractions D$\to\omega\pi$ and D$\to\eta\pi$.}
   \label{tab:omegaPi:results}
\end{table}

%%%%%%%%%%%%%%%%%%%%%%%%%%%%%%%%%%%%%%%%%%%%%%%%%%%%%%%%%%%%%%%%%%%%%%%%%%%%%
\section{Branching-fraction \DKsKK}
%%%%%%%%%%%%%%%%%%%%%%%%%%%%%%%%%%%%%%%%%%%%%%%%%%%%%%%%%%%%%%%%%%%%%%%%%%%%%
A \babar measurement\cite{Aubert:2005sm} is the basis of the current PDG\cite{Agashe:2014kda} value:
\begin{align}
   \Gamma(\DKsKK)/\Gamma=\SI{4.47(34)}{\timesten\tothe{-3}}
   \label{eqn:dkskk:pdg}
\end{align}
Since the decay was measurement in the reaction $D^*\to\Dz\pi^\pm$ only a relative normalization is possible (in that case relative to $\KS\pi^+\pi^-$), the precision is only \SI{7.6}{\percent}.

With the large statistic sample at \bes of $\psiprpr\to\DD$ we can measure the branching fraction of the decay with absolute normalization, which in turn reduces the uncertainty. Furthermore an analysis of \DKsKK Dalitz plot is ongoing.
%-----------------------------------------------------------------------------
\subsection{Reconstruction and selection}
%-----------------------------------------------------------------------------
\begin{wrapfloat}{figure}[33]{r}{0.5\textwidth}
   \vspace{-0.8cm}
   \centering
   \includegraphics[width=\textwidth]{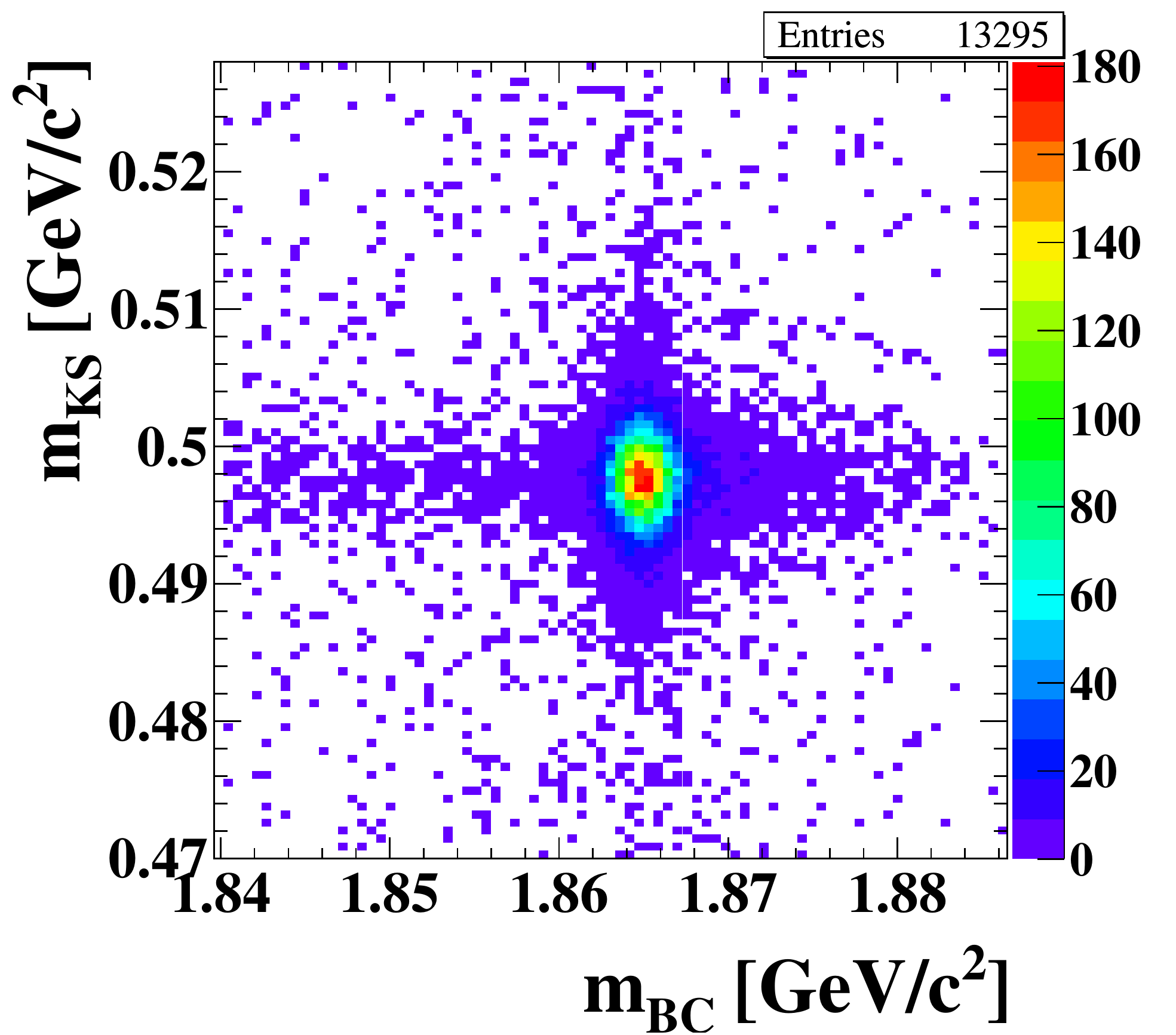}
   \caption{Selected candidates.}
   %\vspace{+1.5cm}
   \label{fig:dkskk:signalPlane}
   \begin{tikzpicture}[overlay]
	   %\draw[help lines] (-5,-4) grid (10,10);
	  \node[rotate=25] at (.5,5.5) (n1) {\color{red}BESIII preliminary};
   \end{tikzpicture}

   \subfloat{
	  \includegraphics[width=\textwidth]{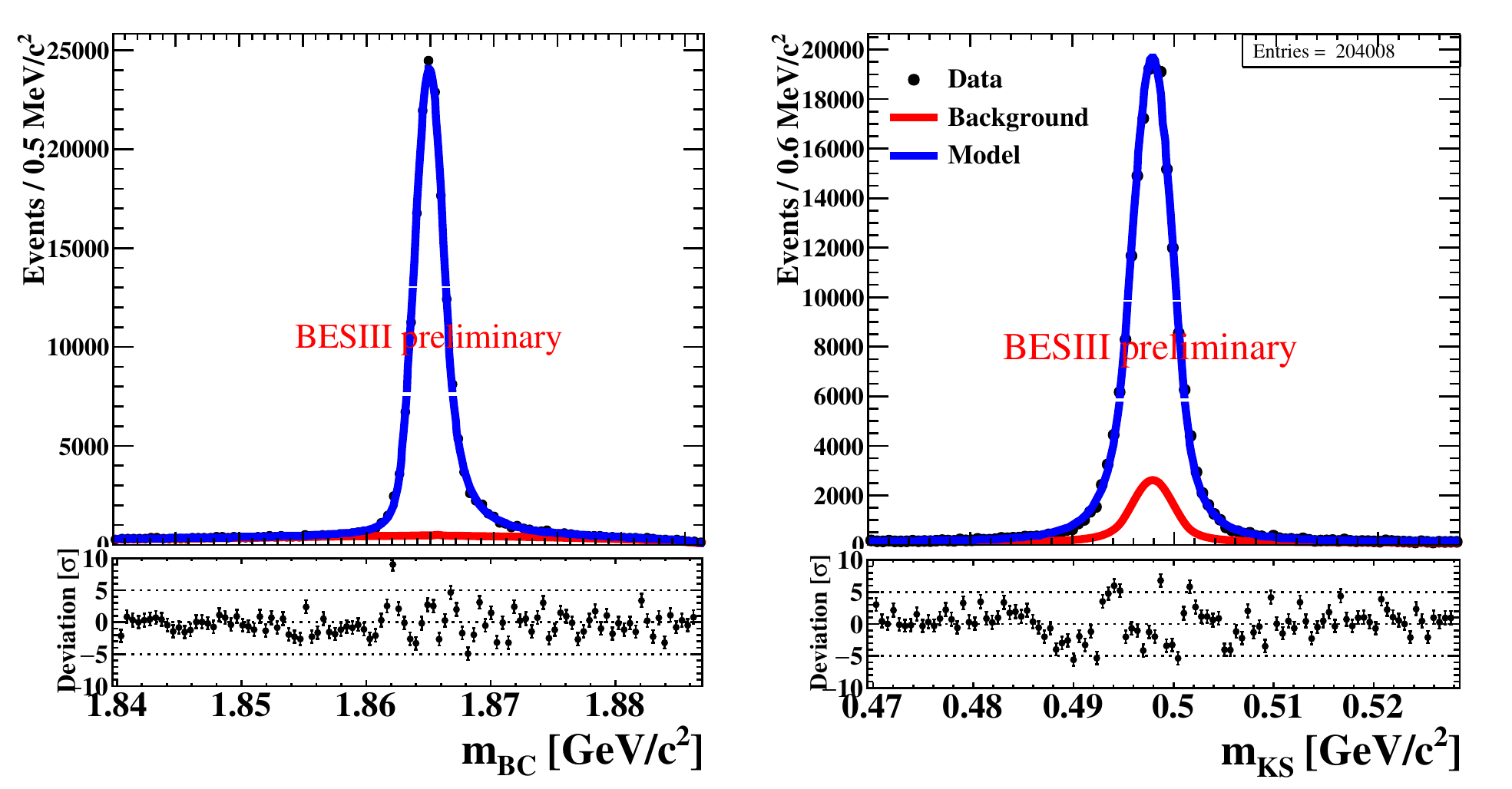}
   }
   \qquad
   \subfloat{
	  \includegraphics[width=\textwidth]{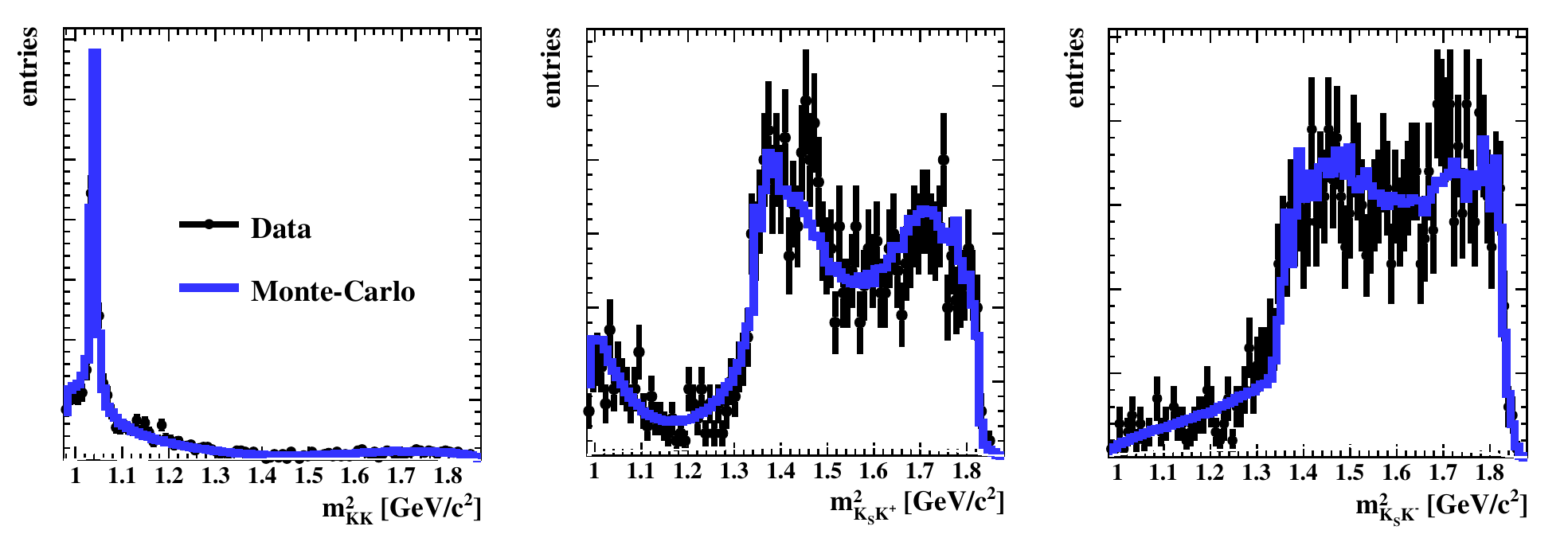}
   }
   \caption{Projections of fit to inclusive MC sample(top) and Dalitz plot projections of signal model(bottom).}
   \label{fig:dkskk:efficiency}
   \begin{tikzpicture}[overlay]
	   %\draw[help lines] (-5,-4) grid (10,10);
	  \node[rotate=25] at (0.5,3.5) (n1) {\color{red}BESIII preliminary};
   \end{tikzpicture}
\end{wrapfloat}

Due to the quantum-correlation of \Dz and \Dzb a branching fraction measurement using the double tag method is very difficult. It would require knowledge of the mixing parameters and the ratio of DCS to CF decays for all tag channels. Therefore we reconstruct the signal decay untagged.

The \KS is reconstructed in the channel $\KS\to\pi^+\pi^-$ and our final state is therefore $K^+K^-\pi^+\pi^-$. We require that the kaon tracks come from the interaction point and pass criteria for particle identification. The \KS candidate is furthermore required to have a significant flight distance. All tracks are fitted with the constraint to make the \Dz mass.

%\begin{wrapfigure}[15]{r}{0.5\textwidth}
   %\vspace{-2cm}
   %\subfloat{
	  %\includegraphics[width=\textwidth]{DKsKK/efficiencyFit}
   %}
   %\qquad
   %\subfloat{
	  %\includegraphics[width=\textwidth]{DKsKK/invmassFit-inputModel}
   %}
   %\caption{Projections of fit to inclusive MC sample(top) and Dalitz plot projections of signal model(bottom).}
   %\label{fig:dkskk:efficiency}
   %\begin{tikzpicture}[overlay]
	   %%\draw[help lines] (-5,-4) grid (10,10);
	  %\node[rotate=25] at (1.0,1.5) (n1) {\color{red}BESIII preliminary};
   %\end{tikzpicture}
%\end{wrapfigure}
The distribution in \KS mass and beam-constraint mass \mBC for all selected signal candidates is shown in Fig.\ref{fig:dkskk:signalPlane}. 
We determine the signal yield by a 2D fit in \KS mass and beam-constraint mass \mBC. According to a simulation study the background consists mainly of \qqbar events.

%-----------------------------------------------------------------------------
\subsection{Efficiency}
%-----------------------------------------------------------------------------
The efficiency of reconstruction and selection is obtained on a inclusive MC sample by the same fitting procedure as on data. This ensures that potential biases cancel in the branching fraction ratio. The projection of the MC sample and the fitted model is shown in Fig.\ref{fig:dkskk:efficiency}. We obtain a value of \SI{0.1719(4)}\xspace. The efficiency is not constant over whole phase space, which leads to a dependence on the MC amplitude model. However our signal amplitude model is in adequate agreement with data so that we can neglect this source of systematic uncertainty.

%-----------------------------------------------------------------------------
\subsection{Systematics and results}
\begin{figure}[tbp]
   \begin{floatrow}
	  \floatbox[]{table}[0.3\textwidth]
	  {%
		 \resizebox{0.3\textwidth}{!}{

			\begin{tabular}{c|c}
			   \toprule
			   \multicolumn{2}{c}{Systematic uncertainties [\%]}\\
			   \midrule
			   PDF shape 				&0.20\\
			   selection 				&0.80\\
			   \midrule
			   \multicolumn{2}{c}{Efficiency}\\
			   \midrule
			   statistics 			&0.33\\
			   PID ($K^+K^-$)			&2.00\\
			   tracking 				&2.00\\
			   \KS \ reconstruction 	&1.50\\
			   \midrule
			   \multicolumn{2}{c}{External}\\
			   \midrule
			   Luminosity measurement	&1.00\\
			   cross-section $\epem\rightarrow\Dzero\DzeroBar$ &1.83\\
			   \KS\ BF 					&0.07\\
			   \midrule
			   Total 					&3.92\\
			   \bottomrule
			\end{tabular}
		 }
		 \vspace{-0.5cm}
	  }
	  {%
		 \caption{Systematic uncertainties.}
		 \label{tab:dkskk:sys}
	  }
	  \floatbox[]{figure}[0.7\textwidth]{%
		 \resizebox{0.75\textwidth}{!}{
			\includegraphics[width=\textwidth]{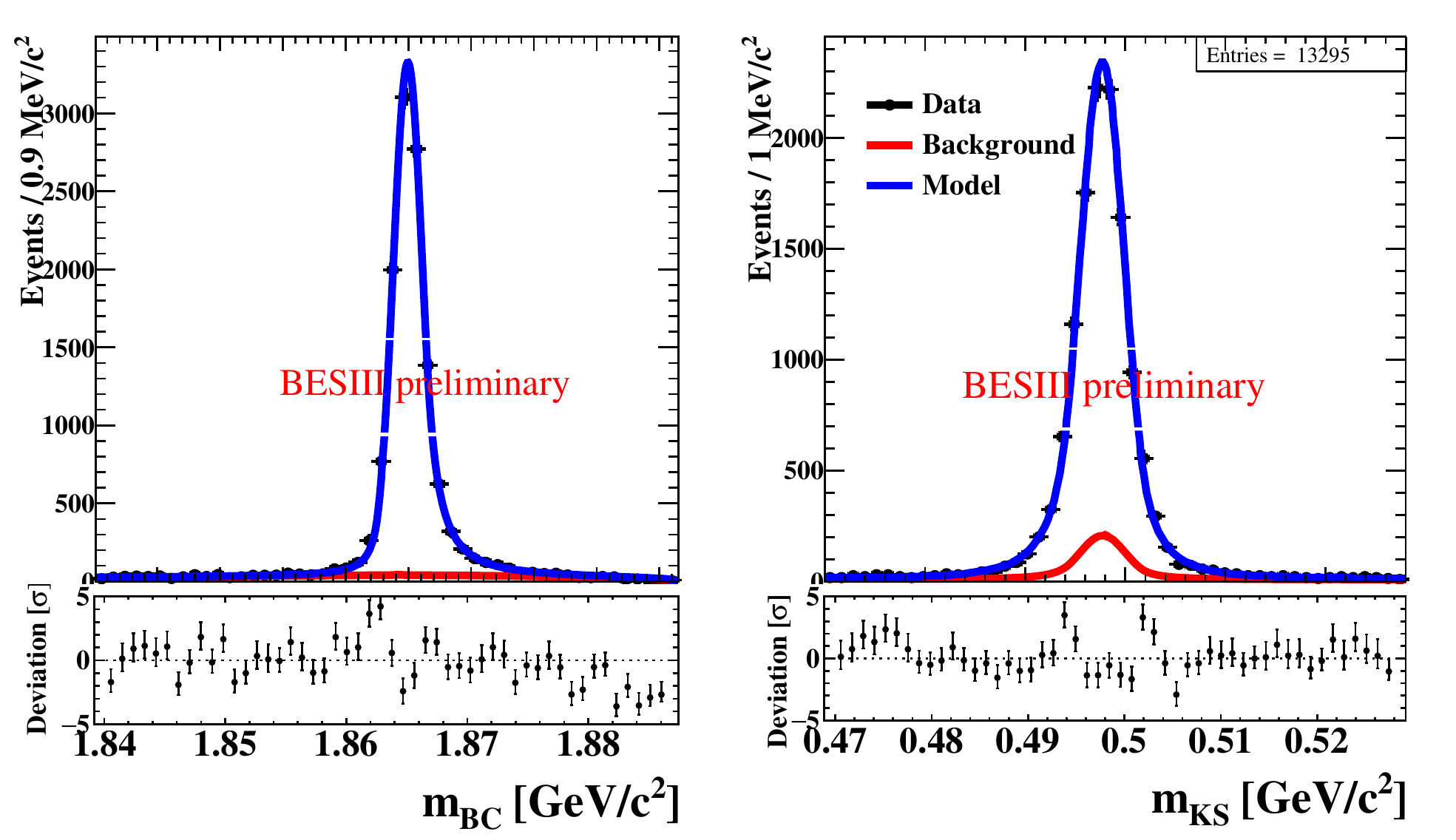}
		 }
		 \vspace{-0.5cm}
	  }{%
		 \caption{Projections of fit model and data sample.}
		 \label{fig:dkskk:dataBF}
	  }
   \end{floatrow}
\end{figure}

%-----------------------------------------------------------------------------
The systematic uncertainties on the branching fraction are listed in Tab.\ref{tab:dkskk:sys}. The largest contributions arise from charged track reconstruction and identification of K$^\pm$ and from the uncertainty of the cross-section measurement $\epem\to\DzDzb$. The total systematic uncertainty is below \SI{4}{\percent}.

The branching fraction can be calculated by:
\begin{align}
   \BR_{\DKsKK} = \frac{N^{sig}}{\epsilon_{BF}\cdot\BR_{\KS\to\pi\pi}\cdot {\lumi}\cdot 2\sigma_{\DzDzb}}
   \label{eqn:dkskk:bf}
\end{align}
The cross-section $\epem\to\Dz\Dzb$ measured by CLEO-c\cite{Dobbs:2007ab} is \SI{3.66(7)}{\nano\barn} and for the branching fraction \KS$\to\pi^+\pi^-$ the PDG\cite{Agashe:2014kda} average is used:

Our preliminary result for the branching fraction \DKsKK is:
\begin{align}
   BF_{data}(\DKsKK) =& \SIerrs{4.622}{0.045}{0.181}{\timesten\tothe{-3}} \\
   \label{eqn:dkskk:bfData}
\end{align}
The total uncertainty is \SI{4}{\percent} which is an improvement of the PDG value by almost a factor of 2. The agreement with the PDG value is better 1$\sigma$.

%%%%%%%%%%%%%%%%%%%%%%%%%%%%%%%%%%%%%%%%%%%%%%%%%%%%%%%%%%%%%%%%%%%%%%%%%%%%%
\section{Summary}
%%%%%%%%%%%%%%%%%%%%%%%%%%%%%%%%%%%%%%%%%%%%%%%%%%%%%%%%%%%%%%%%%%%%%%%%%%%%%
We present preliminary results from studies of hadronic charm decays. We present the first observation of the decay $\Dp\to\omega\pi^+$ with a branching fraction of  \SIerrs{2.74}{0.58}{0.17}{\timesten\tothe{-4}} and find evidence for the decay $\Dz\to\omega\pi^0$ with a branching fraction of \SIerrs{1.05}{0.41}{0.09}{\timesten\tothe{-4}}. Furthermore we measured the branching fractions $D^{(+,0)}\to\eta\pi^{(+,0)}$ in good agreement with the PDG average.

The decay \DKsKK is studied using an untagged method and a preliminary branching fraction of \SIerrs{4.622}{0.045}{0.181}{\timesten\tothe{-3}} is obtained. The measurement is the first absolute measurement and reduces the uncertainty of this branching fraction by almost a factor of 2. An analysis of the Dalitz plot is currently ongoing.
%%%%%%%%%%%%%%%%%%%%%%%%%%%%%%%%%%%%%%%%%%%%%%%%%%%%%%%%%%%%%%%%%%%%%%%%%%%%%
\printbibliography

\end{document}